\documentclass{emulateapj}
\usepackage{apjfonts}
\usepackage{graphicx}

\def\d3{$\delta_{3}$ }
\def\1d3{$(1 + \delta_{3})$ }
\def\l1d3{$\log_{10}(1 + \delta_{3})$ }

\def\s3{$\Sigma_{3}$}

\def\24m{24 $\mu$m}
\def\um{$\mu$m }

\def\sm{$M_{*}$}

\def\mpc{$h^{-1}$ Mpc}
\def\kms{${\rm km~s^{-1}}$ }

\def\Msolar{$\rm M_{\odot}$}

\def\Aw{$A_{\omega}$}

\shorttitle{Clustering properties for $\textit{B\lowercase{z}K}$-selected galaxies at \lowercase{z} $\sim$ 2 in GOODS-N}
\shortauthors{Lin et al.}

\begin{document}

\title{Clustering properties of $\textit{B\lowercase{z}K}$-selected galaxies in GOODS-N: Environmental quenching and triggering of star formation at $z \sim 2$}

\author{Lihwai Lin \altaffilmark{1}, Mark Dickinson \altaffilmark{2}, Hung-Yu Jian \altaffilmark{3},  A. I. Merson \altaffilmark{4}, C. M.  Baugh \altaffilmark{4}, Douglas Scott \altaffilmark{5}, Sebastien Foucaud \altaffilmark{6}, Wei-Hao Wang \altaffilmark{1}, Chi-Hung Yan \altaffilmark{1}, Hao-Jing Yan \altaffilmark{7}, Yi-Wen Cheng \altaffilmark{8}, Yicheng Guo \altaffilmark{9},  John Helly \altaffilmark{4}, Franz Kirsten \altaffilmark{10}, David C. Koo \altaffilmark{11}, Claudia del P. Lagos \altaffilmark{4}, Nicole Meger \altaffilmark{5}, Hugo Messias \altaffilmark{12}, Alexandra Pope \altaffilmark{9}, Luc Simard \altaffilmark{13,14}, Norman A. Grogin \altaffilmark{15}, Shiang-Yu Wang \altaffilmark{1}}

\altaffiltext{1}{Institute of Astronomy \& Astrophysics, Academia Sinica, Taipei 106, Taiwan   (R.O.C.); Email: lihwailin@asiaa.sinica.edu.tw}
\altaffiltext{2}{National Optical Astronomy Observatory, 950 N. Cherry Ave., Tucson, AZ 85719, USA}
\altaffiltext{3}{Department of Physics, National Taiwan University, Taipei, Taiwan (R.O.C.)}
\altaffiltext{4}{Institute for Computational Cosmology, Department of Physics, Durham University, South Road, Durham DH1 3LE, UK}
\altaffiltext{5}{Department of Physics \& Astronomy, University of British Columbia, 6224 Agricultural Road, Vancouver, BC V6T 1Z1, Canada}
\altaffiltext{6}{Department of Earth Sciences, National Taiwan Normal University, N$^{\circ}$88, Tingzhou Road, Sec. 4, Taipei 11677, Taiwan (R.O.C.)}
\altaffiltext{7}{Center for Cosmology and AstroParticle Physics, The Ohio State University, 191 West Woodruff Avenue, Columbus, OH 43210, USA}
\altaffiltext{8}{Institute of Astronomy, National Central Universe, Jhongli, Taiwan (R.O.C.)}
\altaffiltext{9}{Astronomy Department, University of Massachusetts, 710 N. Pleasant St., Amherst, MA 01003, USA}
\altaffiltext{10}{Argelander-Institut f\"{u}r Astronomie, University of Bonn, Germany}
\altaffiltext{11}{UCO/Lick Observatory, Department of Astronomy and Astrophysics, University of California, Santa Cruz, CA 95064, USA}
\altaffiltext{12}{Centro de Astronomia e Astrof\'{i}sica da Universidade de Lisboa, Observat\'orio Astron\'omico de Lisboa, Tapada da Ajuda, 1349-018 Lisboa, Portugal.}
\altaffiltext{13}{National Research Council of Canada, Herzberg Institute of Astrophysics, 5071 West Saanich Road, Victoria, British Columbia, Canada}
\altaffiltext{14}{Department of Physics and Astronomy, University of Victoria, Victoria, British Columbina V8P 1A1, Canada}
\altaffiltext{15}{Space Telescope Science Institute, 3700 San Martin Drive, Baltimore, MD 21218, USA}

\begin{abstract}
Using a sample of $\textit{BzK}$-selected galaxies at $z \sim 2$ identified from the CFHT/WIRCAM near-infrared survey of GOODS-North, we discuss the relation between star formation rate (SFR), specific star formation rate (SSFR), and stellar mass (\sm), and the clustering of galaxies as a function of these parameters.  For star-forming galaxies (\textit{sBzK}s), the UV-based SFR, corrected for extinction, scales with the stellar mass as SFR $\propto$ \sm$^{\alpha}$ with $\alpha = 0.74\pm0.20$ down to \sm$~$ $\sim 10^{9}$ \Msolar, indicating a weak dependence on the stellar mass of the star formation rate efficiency, namely, SSFR. We also measure the angular correlation function and hence infer the correlation length for \textit{sBzK} galaxies as a function of \sm, SFR, and SSFR, as well as $K$-band apparent magnitude. We show that passive galaxies (\textit{pBzK}s) are more strongly clustered than \textit{sBzK} galaxies at a given stellar mass, mirroring the color$-$density relation seen at lower redshifts.
We also find that the correlation length of \textit{sBzK} galaxies ranges from 4 to 20 \mpc, being a strong function of $M_{K}$, \sm, and SFR.
On the other hand, the clustering dependence on SSFR changes abruptly at $2\times 10^{-9}$ yr$^{-1}$, which is the typical value for ``main sequence'' star-forming galaxies at $z \sim 2$. We show that the correlation length reaches a minimum at this characteristic value, and is larger for galaxies with both smaller and larger SSFRs; a dichotomy that is only marginally implied from the predictions of the semi-analytical models. Our results suggest that there are two types of environmental effects at work at $z \sim 2$. Stronger clustering for relatively quiescent galaxies implies that the environment has started to play a role in quenching star formation.  At the same time, stronger clustering for galaxies with elevated SSFRs (``starbursts'') might be attributed to an increased efficiency for galaxy interactions and mergers in dense environments.

\end{abstract}


\section{INTRODUCTION}

When exactly galaxies formed, and how their evolutionary histories are associated with their environment, are among the most important problems in extra-galactic astronomy. Recent studies have suggested that the critical epoch for building up galaxy mass and shaping galaxy properties is around $z \sim 2$, as the peak in the cosmic star formation rate density lies
at this redshift \citep{dic03,hop04,hop06,cha10}. Similar trends have also been discovered for AGN/QSO activity \citep{ric06}. Furthermore, the number and stellar mass density of quiescent galaxies have been found to build up rapidly since this epoch \citep{arn07}. Therefore, it is essential to probe the properties and abundance of galaxies at this epoch and beyond in order to understand galaxy evolution. One method which selects $z > 1$ galaxies is the so-called \textit{BzK} color selection. Specifically at $1.4 < z < 2.5$, this method has been shown to efficiently separate star-formation dominated sources from those already in a passive phase of galaxy evolution \citep{dad04}.

While large numbers of $z \sim 2$ galaxies have been routinely discovered by deep imaging surveys, we still lack a complete picture of the connection between different galaxy properties at this era, and in particular the interplay between the host dark matter halos and galaxies. One method to quantify the masses of dark matter halos that host the galaxies is to measure the amplitude of galaxy clustering. Within the cold dark matter (CDM) model, it is well known that more massive dark matter haloes are more strongly clustered \citep[e.g.,][]{bau99,mo02}. Despite this, the details of the spatial distribution of galaxies may be complex due to the additional physical processes affecting galaxies during their formation and evolutionary histories, for instance, super-novae
and AGN feedback \citep[e.g.,][]{kay02,ben03,som08}. However, under simple assumptions or through halo occupation distribution (HOD) modeling, one can infer the host halo mass of galaxies through their clustering strength \citep{ber02,kra04,zhe05,tin05}.

Observations at $z < 1.5$ suggested that the clustering amplitude depends strongly on galaxy properties such as morphology, color, and luminosity \citep{nor02,zeh02,coi08,del11}. At $z \sim 3$ and higher redshifts it has also been found that the clustering strength increases with UV-continuum luminosity and hence SFR for UV-selected Lyman Break Galaxies \citep{gia01,fou03,ade05,kas06,lee06,ouc04,yos08}. However, the situation for $z \sim 2$ populations is less clear. There have been several attempts to study the clustering of \textit{BzK}-selected populations. Most of these focused on the dependence of clustering on $K-$band magnitude and galaxy types \citep[star-forming \textit{BzK}s vs passive \textit{BzK}s;][]{kon06,hay07,bla08,har08,mcc10}. Only a few studies have probed the clustering dependence on other properties such as stellar mass (\sm), star formation rate (SFR), and specific star formation rate (SSFR = SFR/stellar mass) at $z \sim 2$ \citep{fou10,wak11,mag11,sav11}. This was partly due to the difficulty in obtaining deep and yet wide-field near-infrared (NIR) imaging, which is critical for the $\textit{BzK}$ selection. The samples were typically drawn from either small deep fields or shallow wide surveys, and therefore lacked the dynamic range required for sampling the properties of interest.

Thanks to the new-generation of wide-field NIR imagers, deep NIR galaxy surveys have now become possible, allowing one to probe both bright and faint populations of galaxies simultaneously. In this work, we carry out the first systematic study of clustering as a function of various galaxy properties for $\textit{BzK}$-selected galaxies down to $K_{\rm s} \sim 24.0$ (AB mag), selected from the multiwavelength data in the Great Observatories Origins Deep Survey \citep[GOODS:][]{gia04} North field (GOODS-N). The relation between clustering strength and SSFR is particularly interesting because we can gain insight into the relation between the halo mass, and thereby environment, and star formation efficiency. Previous work suggests that environment plays an important role at low redshifts in quenching the star formation of galaxies \citep{dre80,bal98,gom03,kau04}, but that this effect reverses at $z \sim 1$ \citep{elb07,coo08}. However, a recent study by \citet{qua12} investigating the quiescent fraction as a function of local density found in contrast that the SFR$-$density relation continues at least out to $z \sim 1.8$. The role of environment at $z > 1$ is thus still under debate, although different results may be attributed to different definitions of the 'SFR$-$density' relation (see \S 4.2). Instead of using local density measurements, in this work we use the clustering strength as a probe of halo mass and thus environment. This has the advantage of avoiding noisy density measurements due to either photometric redshift errors, or incomplete sampling in the spectroscopic redshift samples given that the method relies on the projected distributions alone.

The paper is organized as follow. The data sets, $BzK$ samples, and the methods used in this paper to compute the stellar mass and star formation rate are described in \S 2. The results of the relation between SFR, SSFR, and \sm, and the clustering analysis are presented in \S 3. We discuss the implications of our results in \S4, followed by a brief summary in \S 5.

Throughout this paper we adopt the following cosmology: \textit{H}$_0$ = 100$h$~\kms Mpc$^{-1}$, $\Omega_{\rm m} =
0.3$ and $\Omega_{\Lambda } = 0.7$. We adopt the Hubble constant $h$ = 0.7 when calculating rest-frame
magnitudes. We use a Salpeter IMF when deriving stellar masses and star formation rates. All magnitudes are given in the AB system.

\section{DATA, SAMPLE SELECTIONS, AND METHODS \label{sec:data}}
GOODS-North is one of the most heavily studied extra-galactic fields, with rich multiwavelength data sets. The NIR observations in GOODS-N were carried out with the
Wide-Field Near Infrared Camera (WIRCAM) on the CFHT during 2006$-$2009. These include 27.4 hr of integration in $J$ band obtained by a
Taiwanese program (PI: L. Lin) and 31.9 hr of integration in $K_s$ band obtained by
Hawaiian (PI: L. Cowie) and Canadian (PI: L. Simard) programs. Results from the WIRCAM $K_{\rm s}$-band imaging were recently published by \citet{wan10}; here we use our own
reductions (L. Lin et al., in preparation). The data were first pre-processed using
the SIMPLE Imaging and Mosaicking PipeLinE \citep{wan10}, and then
combined to produce deep stacks with the AstrOmatic software 'SCAMP' \citep{ber06} and
'SWarp' \citep{ber02} \footnote{http://www.astromatic.net/}. The resulting 5$\sigma$
limiting magnitudes using 2-arcsec diameter circular apertures, reach $J$ =
24.6 and $K_{\rm s}$ = 24.0 in the central 420 arcmin$^{2}$. The optical data used in this work comes from two sources: one is the \textit{HST}/ACS $B_{435}$,
$V_{606}$, $i_{775}$, and $z_{850}$ v2.0 catalog from the GOODS \textit{HST} Treasury Program \citep{gia04}; the other includes the ground-based $U$ band imaging obtained on the KPNO Mayall 4-m telescope and the $BVRiz$ bands taken with Subaru \citep{cap04}. The GOODS-N has also been imaged with IRAC at 3.6, 4.5, 5.8, and 8.0\um , taken as part of the GOODS $Spitzer$ Legacy program (PI: M. Dickinson). The area used in this work is limited to the ACS coverage of 10' $\times$ 16' = 160 arcmin$^{2}$.

The photometry for each band is done using a software package with object template-fitting method \citep[TFIT;][]{lai07} based on the ACS $z$-band detections. For
each object, TFIT constructs a template using the spatial position and morphology of
the object in the ACS $z$-band image. Such a template is then convolved with the PSF of other low-resolution images and then fit to the images of the object in other bands.
The best-fitting fluxes are then considered as the final fluxes of the
object in low-resolution bands. The ACS photometry is measured in dual-mode with SExtractor while photometry in other bands is measured by TFIT with $z$- band template
(N. Grogin, et al. in preparation.)

Our star-forming \textit{BzK} galaxies (hereafter \textit{sBzK}s) are selected using $HST$/ACS $B$ and $z$ bands, and the CFHT WIRCAM $K_{\rm s}$ band, following the \textit{BzK} method developed by \citet{dad04}. We add 0.3 mag to the $z - K$ color of the dividing lines that separate the \textit{sBzK} galaxies, passive \textit{BzK} galaxies (hereafter \textit{pBzK}s), stars, and foreground/background galaxies (hereafter non-\textit{BzK}s), in order to account for the photometry offsets due to different filter systems and zeropoint calibrations. The \textit{sBzK} galaxies are selected to be those objects with
\begin{equation}\label{eq_sbzk}
(z - K) - (B - z) > 0.1,
\end{equation}
while \textit{pBzK} galaxies are defined as those objects with
\begin{equation}\label{eq_pbzk}
(z - K) - (B - z) < 0.1 \cap (z - K) > 2.8.
\end{equation}
Stars are also identified by the relation
\begin{equation}\label{eq_star}
(z - K) - (B - z) < -0.2 + (B - z) \times 0.3.
\end{equation}

\begin{figure}
\includegraphics[angle=-270,width=12cm]{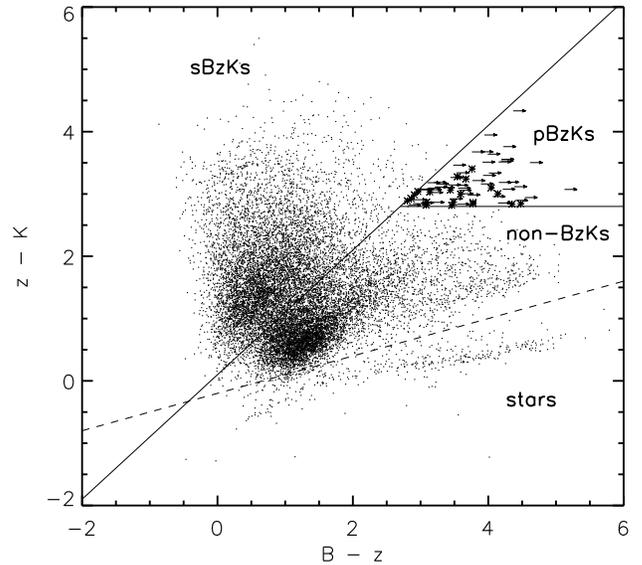}

\caption{$B - z$ vs. $z - K$ diagram for all objects in GOODS-N. The \textit{sBzK}s, \textit{pBzK}s, foreground/background galaxies (non-\textit{BzK}s), and stars are classified according to the four regions separated by the solid and dashed lines as adopted by \citet{dad04} with additional small adjustments to account for the filter differences. Stars represent the \textit{pBzK} galaxies that are detected in the $B$-band while the arrows denote the objects not detected in the $B$-band, but their $B_{2\sigma lim} - z$ and $z - K$ colors satisfy the p\textit{BzK} criteria.
\label{fig:bzk}}
\end{figure}

The $B - z$ versus $z - K$ distribution of our sample is shown in Figure \ref{fig:bzk}.
We require signal-to-noise ratios $S/N > 5$ in both the $z$ and $K$ bands \footnote{The typical magnitudes with $S/N = 5$ in our sample are 27.9 in the $B$ band, 27.3 in the $z$ band, and 24.0 in the $K_{s}$ band.}. Moreover, we further apply a magnitude cut of [4.5] $<$ 25.3 mag, corresponding to a 5-$\sigma$ limiting mag in [4.5] so we can obtain reliable stellar mass measurement. For objects undetected in the $B$-band, if their $B_{2\sigma lim} - z$ and $z - K$ colors satisfy the \textit{pBzK} criteria, we classify them as \textit{pBzK} galaxies, where $B_{2\sigma lim}$ is the $B$ band $2\sigma$ limiting magnitude. On the other hand, we treat the remaining objects without detections in $B$ as unclassified \textit{BzK} galaxies (hereafter \textit{uBzK}s). In total, we classified 4496 \textit{sBzK}s, 45 \textit{pBzK}s, 179 \textit{uBzK}s, 341 stars, and 5791 non-\textit{BzK}s. The $\textit{BzK}$ sample used in this work probes galaxies fainter than previous clustering studies of \textit{BzK}s by 0.5-1 mag in the $K$-band.

The photometric redshifts (hereafter $z_{phot}$) were obtained by fitting ground-based $UBVRizJK$  photometry using the public ``BPZ'' photometric redshift code \citep{ben00}. The rms of our $z_{phot}$ when compared to the spectroscopic redshifts of the same objects is about $0.07(1+z)$ with 6.5\% outliers at $0.1 < z < 3.0$. Spectroscopic redshifts used for our $z_{phot}$ calibration were compiled from several published and unpublished sources, particularly \citet{coh00}, \citet{wir04}, \citet{red06b}, \citet{bar08}, and D. Stern et al. (in preparation). This spectroscopic redshift catalog consists of  $4308$ objects up to $z \sim 6.7$, among which 639 objects satisfy our \textit{sBzK} selection. Although the spectroscopic sample contains objects with $K-$band mag down to $~25.0$, they are systematically brighter than the full $\textit{BzK}$ sample used in this work. The  $K-$band magnitude of the 639 \textit{sBzK}-selected galaxies with spectroscopic redshift measurement peaks around 23.0, while that of the full \textit{sBzK} sample peaks at $K \sim$ 24.0.

\begin{figure}
\includegraphics[angle=-270,width=12cm]{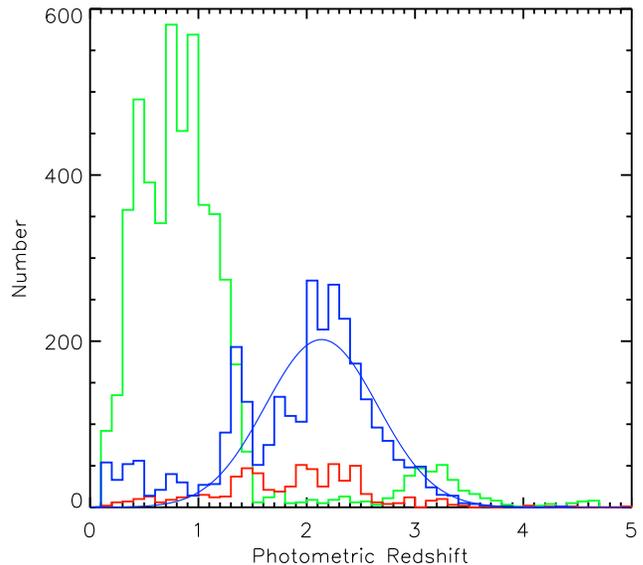}

\caption{Photometric redshift distributions for \textit{sBzK} (blue histogram) and non-\textit{BzK} galaxies (green histogram). The blue curve represents the best fit of the Gaussian parameterizing the $z_{phot}$ distribution for \textit{sBzK} galaxies. Objects with photometric redshift value smaller than 0.1 are not shown. For comparison, we also show the spectroscopic redshift distribution for those \textit{sBzK} galaxies that have a spectroscopic redshift identification (red histogram).
\label{fig:zp}}
\end{figure}

The $z_{phot}$ distributions of the \textit{sBzK} and the non-\textit{BzK} populations, together with the spectroscopic redshift distribution of \textit{sBzK}-selected galaxies which have a spectroscopic identification, are shown in Figure \ref{fig:zp}. It can be seen that although the original design of the \textit{sBzK} selection is to pick up galaxies at $1.4 < z < 2.5$, the redshift distribution of our sample suggests that there is non-negligible contamination from galaxies at $z< 1.4$ as well as those at $z > 2.5$ \citep[also see][]{bar08}. Such contamination however does not affect our clustering results in Section \ref{sec:clustering} since we take into account redshift distribution when converting the angular cluster amplitude into the real-space clustering strength. For those \textit{BzK} galaxies which do not have a photometric redshift measurement because of detection in limited bandpasses or whose redshifts fall out of the redshift range $1 < z_{phot} < 4$, we assign their reshifts to be the median $z_{phot}$ for a given subsample used in the clustering analysis. None of our conclusions change if we restrict our sample to only those with photometric redshift measurement.

The ACS $B_{435}$ band samples rest-frame wavelengths ranging from 1200 to 1800 \AA$~$for the \textit{BzK} galaxies ($1.4 < z < 2.5$), allowing us to estimate the rest-frame UV luminosity at 1500 \AA. According to \citet{dad04}, the reddening for \textit{sBzK} galaxies correlates well with the observed ($B - z$) color in the following way:

\begin{equation}\label{eq_extin}
E(B - V) = 0.25(B - z + 0.1)_{AB}.
\end{equation}
One potential caveat of this approach is that the above empirical relation is derived based on a brighter \textit{sBzK} sample with $K < 22$ \citep{dad04}, and has not been directly tested for fainter $K$--selected galaxies. However, the color-reddening relation for \textit{sBzK} galaxies
was defined  assuming the dust attenuation law from \citet{cal00}, and mid- and far-infrared observations \citep{red06a,red12} have shown that this law works well, on average, for typical $L^\ast$ Lyman break galaxies at $z \sim 2$, which significantly overlap with fainter \textit{sBzK}-selected samples.
Considering the Calzetti extinction law \citep{cal00}, we compute the SFR by converting the extinction corrected $L_{\nu}$(1500 \AA) using the following equation from \citet{dad04}:

\begin{equation}\label{sfr}
\rm{SFR(M_{\odot} yr}^{-1}) = L_{\nu}(1500 \AA)/(8.85\times 10^{20}\rm{WHz}^{-1}).
\end{equation}
To estimate the stellar mass, we follow a similar method to that adopted in \citet{dad04}, but using the observed IRAC 4.5 \um magnitude and [3.6] - [4.5] color instead. We parameterize the stellar mass in the form of :

\begin{equation}\label{eq:sm}
log(M_{*}/M_{\odot}) = 11.0 -0.4([4.5] - a_{1}(z)) + a_{2}(z)([3.6] - [4.5]).
\end{equation}
The $a_{1}$ and $a_{2}$ values given in Table \ref{tab:smfit} are estimated based on a calibration of Eq. \ref{eq:sm} against the stellar mass derived from the SED fitting code 'FAST' \citep{kri09} for a subset of bright \textit{sBzK} galaxies, which have decent detections in multiple bands and spectroscopic redshift measurements. Although the depth of the 4.5 \um data allows us to probe the \textit{BzK} sample down to \sm $\sim$ $10^{9}$ \Msolar, our sample is also limited by the depth in $B-$, $z-$, and $K-$ bands which are used for the \textit{BzK} selection. Similar to Eq. \ref{eq:sm}, one can also estimate stellar mass using the $K-$ band mag and the $z-K$ color as adopted by \citet{dad04}. As a result, we are likely missing galaxies with red $z-K$ colors in the stellar mass range $10^{9.0}$ \Msolar $<$ \sm $<$ $10^{10}$ \Msolar~ given our depth in $K$. In this work, we carry out the analysis down to \sm $\sim$ $10^{9}$ \Msolar, and we will discuss the effect caused by the incompleteness.

\section{RESULTS}
\subsection{SFR $-$ \sm$~$and SSFR $-$ \sm$~$relations \label{sec:sfrssfrsm}}

\begin{figure*}
\includegraphics[angle=-270,width=19cm]{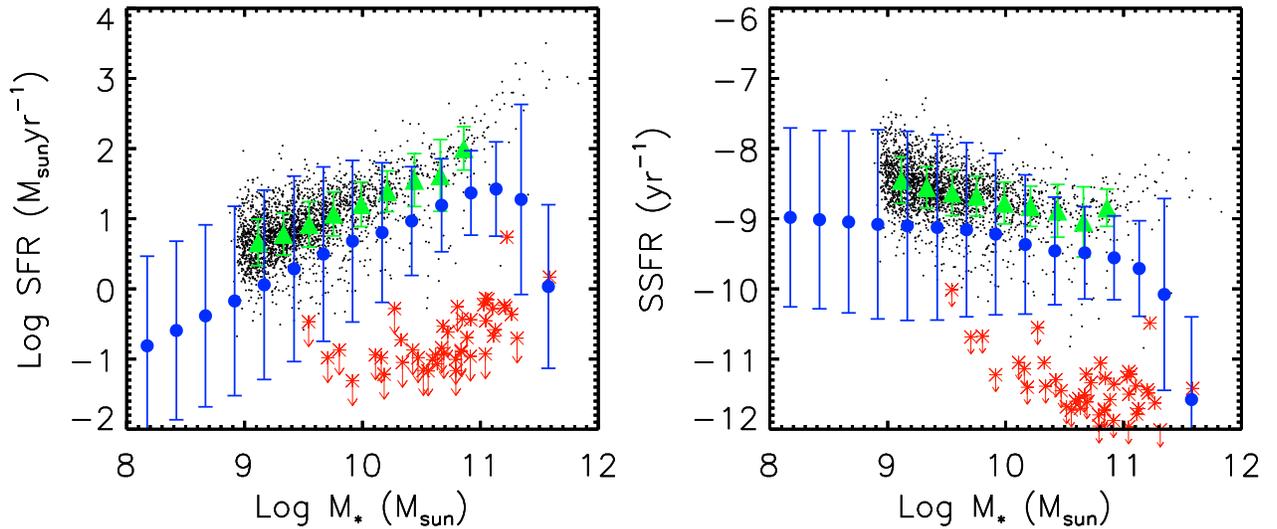}

\caption{Left: SFR and stellar mass correlation for \textit{BzK} galaxies in GOODS-N. Right: SSFR and stellar mass correlation. \textit{sBzK} galaxies are shown using small black dots. The green triangles and blue filled circles represent the median values of the star-forming galaxies in the GOODS-N sample and in the GALFORM model predictions respectively. Star-forming galaxies in the GALFORM model refer to those galaxies with SSFR greater than $10^{-11}$ yr$^{-1}$. \textit{pBzK} galaxies in GOODS-N are shown using red stars. The ones with arrows represent those \textit{pBzK} galaxies that are undetected in $B$. The upper limits of SFR or SSFR are estimated using the 2-$\sigma$ $B$-band limiting magnitudes. The error bars represent the 1-$\sigma$ distribution in each stellar mass bin.
\label{fig:sfrssfr}}
\end{figure*}

We begin by showing SFR versus stellar mass for a subset of the \textit{BzK} sample with 1.8 < $z_{phot}$ < 2.2 in Figure \ref{fig:sfrssfr} (left panel). Although the SFR is estimated using the UV light corrected for dust extinction, rather than using the full SED fitting, there is a clear sequence of \textit{sBzK} galaxies in the SFR$-$\sm $~$relation, as seen in other work that includes IR SFR tracers \citep{wuy11,rod11}. Using SFR $\varpropto$ $M_{*}^{\alpha}$, the slope $\alpha$ for our \textit{sBzK} galaxies is
found to be $0.74\pm0.20$ over the stellar mass range $10^{9}$ \Msolar $<$ \sm $<$ $10^{11}$ \Msolar. Both the slope and the normalization are in good agreement with previous studies at $z \sim 2$ \citep{dad07,pan09,wuy11,rod11}. However, this work extends the stellar mass limit further down to $10^{9}$ \Msolar. As mentioned in \S2, it is possible that we are missing galaxies with red $z-K$ colors, namely, the relatively quiescent galaxies, in the low mass end. If this is true, we would overestimate the SFR for less massive \textit{sBzK} galaxies, and therefore the slope of our derived SFR$-$\sm$~$relation should be regarded as a lower limit.

Also shown in Figure \ref{fig:sfrssfr} is the 1-$\sigma$ distribution of the SFR$-$\sm$~$relation at $z \sim 2$ as predicted by the \citet{lag11} GALFORM model, implemented on the halo merger trees of the Millennium Simulation \citep{spr05}. The GALFORM model is a semi-analytical model of galaxy formation that was originally introduced in \citet[see Baugh 2006 for a review]{col00}. The SFR and \sm$~$ are directly taken from the GALFORM output. The IMF adopted in the model is a Kennicutt IMF \citep{ken83}, so we add 0.3 dex to both the SFR and \sm$~$ to convert them into a Salpeter IMF. It appears that the model predicts larger scatters  in SFR and under-predicts the SFR for a given stellar mass compared to our observations by roughly a factor of 2$-$4.

In the right panel of Figure \ref{fig:sfrssfr} we plot the SSFR as a function of the stellar mass.
Our data suggest that the \textit{sBzK} galaxies are forming stars with slightly weaker dependence on stellar mass than what is observed
at lower redshifts \citep{noe07,cow08}. The slope of our SSFR$-$\sm$~$relation ($-0.26\pm0.20$) is flatter than other work based on optical to mid-IR SFR tracers \citep{feu05b,red06a,erb06,pap06,rod10} at $z \sim 2$, but is consistent with the results derived from radio SFR tracers \citep{dun09,pan09}. This flattening of the SSFR-\sm$~$relation at $z \sim 2$ has recently been interpreted as the dawn of downsizing \citep{pan09}, in the sense that star formation had not ceased in massive galaxies at these redshifts.

It is worth noting that some of the \textit{sBzK} galaxies, in particular for more massive ones, have very low SSFR consistent with that of \textit{pBzK} galaxies, as revealed in Figure \ref{fig:sfrssfr}. This may be partly due to the contamination of passive galaxies that are being scattered into the \textit{sBzK} selection \citep{qua07}, leading to a potential underestimation of the median SSFR for intrinsically massive star-forming galaxies. In addition, the inclusion of passive galaxies in the \textit{sBzK} sample may also affect the results of clustering of galaxies with low SSFR for which we will discuss in \S \ref{sec:clustering}. Nevertheless we expect that the contamination rate should be low given the much lower number density of passive galaxies compared to that of \textit{sBzK} galaxies and hence our main results regarding the SSFR (or SFR) $-$ \sm$~$relation and the clustering analysis should not be significantly affected.

\subsection{Clustering Properties \label{sec:clustering}}

\begin{figure*}
\includegraphics[angle=-270,width=19cm]{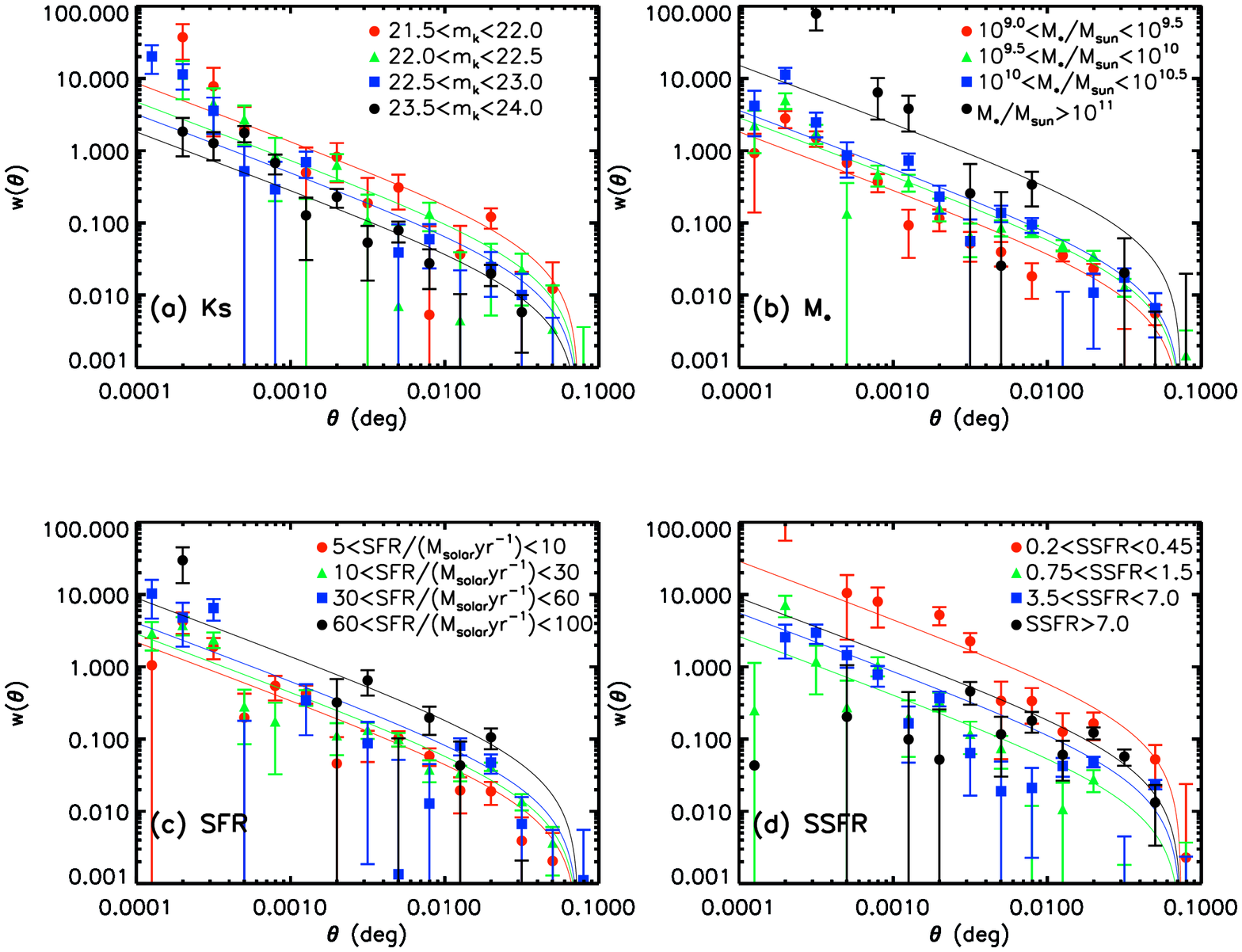}

\caption{The angular correlation function selected in (a) $K$-band magnitude, (b) stellar mass, (c) star formation rate, and (d) specific star formation rate for \textit{sBzK} galaxies. In each panel, the colors represent the measurement of samples selected in different bins. The lines represent the best fit power law using a fixed slope of $\gamma = 1.8$, with the integral constraint incorporated. Bins with $\theta < 0.001$ deg are excluded from the fit in order to reduce the contribution from very close pairs. For clarity, only four selected bins are plotted in each panel. The fitting results for the full subsamples are give in Table \ref{tab:r0}. The values of SSFR which appear in the legend are in units of 10$^{-9}$yr$^{-1}$.
\label{fig:wtheta}}
\end{figure*}

\begin{figure*}
\includegraphics[angle=-270,width=19cm]{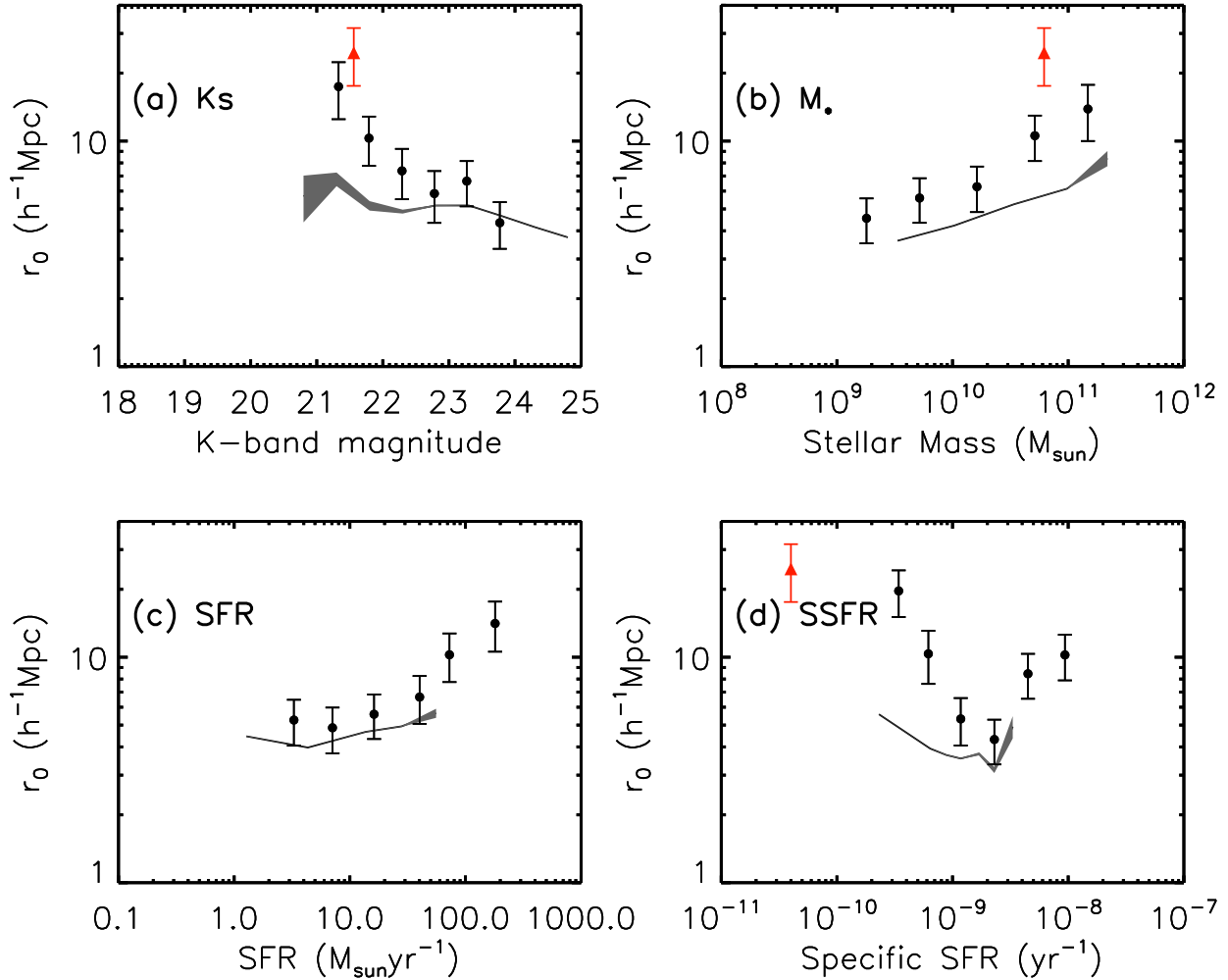}

\caption{Correlation length $r_0$ for \textit{sBzK} galaxies as a function of (a) $K$-band magnitude, (b) stellar mass, (c) star formation rate, and (d) specific star formation rate. The black filled circles denote the results of \textit{sBzK} galaxies in our sample. As a comparison, the results of \textit{pBzK} galaxies are shown in red triangles. The SSFR of \textit{pBzK} galaxies is arbitrary assigned to be 4$\times10^{-11}$ yr$^{-1}$ in the lower-right panel. It can be seen that \textit{pBzK} galaxies are more strongly clustered than most \textit{sBzK} galaxies. The gray shaded areas represent 1-$\sigma$ predictions from the semi-analytical galaxy formation model of \citet{lag11}.
\label{fig:r0}}
\end{figure*}

\begin{figure}
\includegraphics[angle=-270,width=12cm]{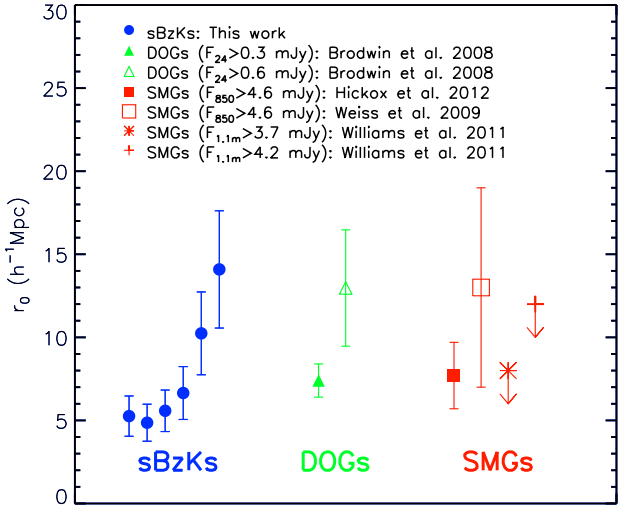}

\caption{Correlation length $r_0$ of \textit{sBzK} galaxies binned in SFR from this study (SFR increases from left to right among blue symbols; see Table \ref{tab:r0}), compared to the measurements of DOGs (green symbols) and SMGs (red symbols) from previous studies.
\label{fig:r0-dusty}}
\end{figure}

\begin{figure}
\includegraphics[angle=-270,width=12cm]{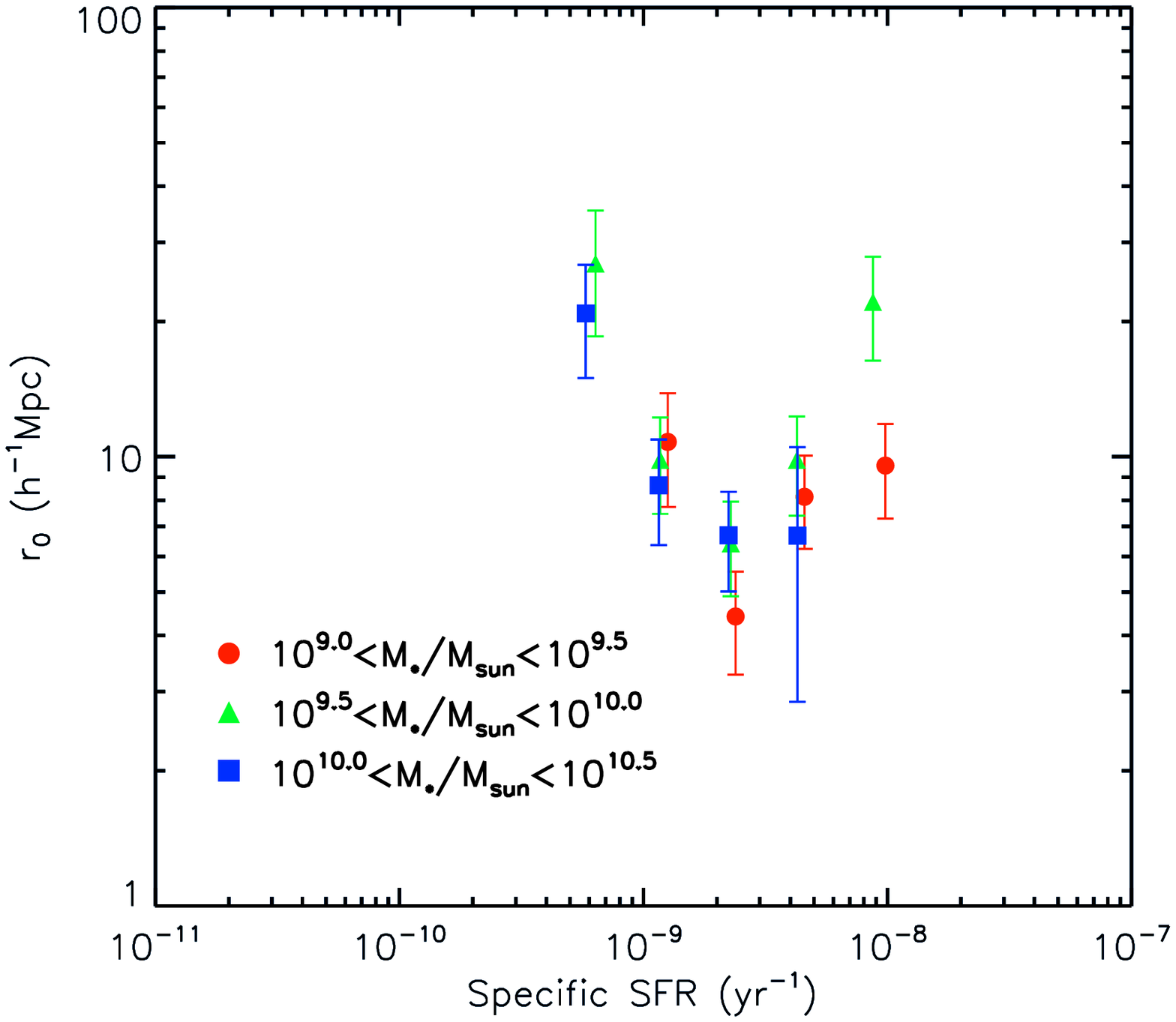}

\caption{The relation between $r_0$ and specific star formation rate of \textit{sBzK} galaxies for different stellar mass bins: $9.0 < log(M_{*}$/\Msolar) $< 9.5$ (red circles), $9.5 < log(M_{*}$/\Msolar) $< 10.0$ (green triangles), and $10.0 < log(M_{*}$/\Msolar) $< 10.5$ (blue squares). It can be seen that the dichotomy shown in the lower-left panel of Figure \ref{fig:r0} still holds at fixed stellar masses.
\label{fig:r0-ssfr}}
\end{figure}

We measure the angular correlation function (ACF) for the \textit{sBzK} galaxies using data and random catalogs with the estimator proposed by \citet{lan93}:
\begin{equation}\label{eq_ac}
\omega(\theta) = [DD(\theta)(n_{R}/n_{D})^{2}-2DR(\theta)(n_{R}/n_{D})+RR(\theta)]/RR(\theta),
\end{equation}
where DD($\theta$), DR($\theta$), and RR($\theta$) are the number of data-data, data-random and random-random pair counts with separations between $\theta$ and $\theta+\delta \theta$, and $n_{D}$ and $n_{R}$ are number of galaxies in the data and random catalogs. Figure \ref{fig:wtheta} shows the ACF measured for the GOODS-N \textit{sBzK} sample, which is binned according to $K$-band magnitudes, \sm, SFR, and SSFR. The error bars in each $\theta$ bin are based on the 1$\sigma$ Poisson statistics of the pair counts in the bins and hence are not correlated across the bins. We fit our ACF with a power law
\begin{equation}
\omega(\theta) = A_{\omega}(\theta^{1-\gamma} - C),
\end{equation}
where $\gamma$ is fixed to be 1.8 and $C$ is the integral constraint which accounts for the finite region of the sky probed in the sample. Following \citet{roc99}, we estimate the integral constraint using
\begin{equation}\label{eq_Cint}
C=\frac{\Sigma RR(\theta)\theta^{-0.8}}{\Sigma RR(\theta)},
\end{equation}
which gives a value of $C = 7.86$ in our case. The 1$\sigma$ error on the amplitude $A_{\omega}$ is computed from the covariance matrix output by the IDL fitting program ``mpfitfun''.

To convert the 2-D clustering amplitude \Aw$~$ into the 3-D correlation length $r_0$, we use Limber's inversion \citep{lim53,mag99}, assuming a Gaussian redshift distribution for our \textit{BzK} sample. Ideally, we should fit a Gaussian to each subsample to derive the peak and the width of its photo-z distribution. However, the sizes of some subsamples are too small to derive meaningful fitting results. Instead, we set the mean of the Gaussian to be the median of the photometric redshift distribution for each subset, and adopt a global width of 0.47 derived from the deconvolution of the photo-z width of the whole s\textit{BzK} sample with the photometric redshift error (see Fig. \ref{fig:zp}). However, it is important to note that $r_0$ is sensitive to the adopted width of redshift distribution, being larger for broader width for a given clustering amplitude \Aw. Therefore if the width varies among different subsamples, it might introduce bias in the trend we see below. To take into account this effect, we also compute $r_0$ in the case of a width of 0.3 which is a good estimate of the minimum value of the width across our subsample\footnote{This also corresponds to the width in the case where approximately 95\% of the \textit{sBzK} galaxies are confined within the redshift range of $1.4 < z < 2.5$.}, and find that $r_0$ decreases by 22\% compared to the case of width = 0.47. We then add this 22\% error in quadrature when estimating the error of $r_0$.

The masses of the halos hosting the galaxies are estimated from $r_0$ in three different ways. The first and simplest method is to assume that each halo hosts only one galaxy, and the selected sample corresponds to a narrow range of halo mass. The second method also assumes one-to-one correspondence between halos and galaxies, but the selected samples are allowed to be hosted by halos above a certain minimum halo mass $M_{\rm min}$. The third method takes into account the Halo Occupation Distribution (HOD), which parameterizes the number of galaxies as a function of halo mass, allowing the possibility that halos can host more than one galaxy. Given our small sample size in each bin, we are not able to perform a full HOD analysis by fitting the ACF with one-halo and two-halo terms. Instead of fitting the HOD parameters, we follow the recipe of \citet{zhe07} and model the number of galaxies hosted by a dark matter halo with mass $M$ as $\langle N(M) \rangle = 1 + M/M_{1}$, with $M_{1} = 20M_{\rm min}$ for $M > M_{min}$, where $M_{min}$ is the minimum halo mass needed to host one galaxy and $M_{1}$ is the mass when a halo hosts two galaxies. We then compute the expected $r_0$ for a given $M_{min}$ and compare that to the $r_0$ derived in our sample. For a given $M_{min}$, one can compute the effective $r_0$ defined as the sum of the $r_0$ of all halos with $M > M_{min}$ weighted by $N(M)$, divided by the total number of halos of with $M > M_{min}$. We note that the HOD adopted here may be oversimplified and not optimized for our sample. Nevertheless, it provides a good approximation to how the halo mass varies for a given $r_0$ due to differences in the $r_0$$-$mass conversions. The derived clustering strength, inferred halo mass from the above three approaches, as well as other sample characteristics, are given in Table \ref{tab:r0}.

Figure \ref{fig:r0} plots the 3-D correlation length as functions of $K_{s}$, \sm, SFR and SSFR for \textit{sBzK} galaxies.
First, we confirm previous results that the clustering amplitude increases as $K$ brightness increases (Figure \ref{fig:wtheta}a and \ref{fig:r0}a).
This correlation between the $K$-band brightness and the clustering amplitude is often interpreted as clustering strength increasing with galaxy stellar mass, as seen at lower redshifts. Since we now have an improved stellar mass estimator from IRAC fluxes (see Eq. \ref{eq:sm}), we are able to probe the stellar mass dependence of the ACF directly (Figure \ref{fig:wtheta}b and \ref{fig:r0}b). It is evident that there exists a strong dependence of the ACF on stellar mass down to the stellar mass limit of $10^{9}$ \Msolar, which is consistent with the findings by \citet{wak11}.
A similar trend is also seen as a function of the UV-derived star formation rate (Figure \ref{fig:wtheta}c and Figure \ref{fig:r0}c), as reported by \citet{sav11}. Moreover in Figure \ref{fig:r0}b, we find that the \textit{pBzK} galaxies are more clustered than \textit{sBzK} galaxies with similar stellar masses.

On the other hand, there is not a monotonic dependence of clustering strength on the specific star formation rate. The correlation length, $r_0$, decreases with increasing SSFR for galaxies with SSFR below  2 $\times$  $10^{-9}$ yr$^{-1}$, but increases with SSFR for galaxies exceeding this threshold.

As a comparison to the HOD predictions, we also show the predictions for how the clustering strength depends upon the $K$-band, \sm, SFR and SSFR values output by the GALFORM model. The GALFORM results, along with 1-$\sigma$ uncertainties, are indicated by the grey shaded region in Figure \ref{fig:r0}. It can be seen that the semi-analytical model predicts similar trends for $K$-band magnitude and \sm$~$ dependence, as seen in the observations, but this dependence is typically weaker in the model.
On the other hand, although the GALFORM model predictions roughly agree with the $r_0$$-$SFR relation for galaxies with SFR $<$ 50 \Msolar yr$^{-1}$, the models have few galaxies with SFRs as large as those we infer for the real $\textit{sBzK}$ galaxies, limiting the range over which we can make the comparison. As a consequence, it is not straightforward to make comparisons in the $r_0$$-$SSFR relation between model and observations due to the narrower dynamical range of SSFR in the model, although there is marginal hint of the up-turn signature in the model predictions. In order to understand whether this discrepancy is due to selection effects, such as color selection, or the physical treatment in the model, it is necessary to carry out an analysis in the model as it was done in this \textit{BzK} galaxy sample. This will be addressed in a forthcoming paper \citep{mer12}.

\section{DISCUSSION}
\subsection{Comparison of $r_0$ between $\textit{sBzK}$ galaxies and other dusty star-forming galaxies at $z \sim 2$}
In this work, we find that for star-forming $\textit{BzK}$ galaxies at $z\sim 2$, the clustering amplitude decreases with the apparent $K-$band magnitude, in good agreement with previous studies \citep{hay07,har08,mcc10}, but our work extends these studies to less luminous populations by almost one magnitude. We also observe a positive correlation between stellar mass and clustering strength for \textit{sBzK} galaxies. The inferred spatial correlation length is $\sim$ 4.5 \mpc$~$ for galaxies with stellar mass around 2 $\times 10^{9}$ \Msolar$~$ and increases to $\sim 13.9$ \mpc$~$ for galaxies of 1.5 $\times 10^{11}$ \Msolar.

Moreover, the clustering amplitude also depends on SFR, being stronger for galaxies with higher SFR. This trend may not be surprising given the good correlation between SFR and \sm. We find that the least actively star-forming systems, with 1.0 $<$ SFR/(\Msolar yr$^{-1}$) $<$ 5.0 have $r_0$ $\sim$ 5.3 \mpc, suggesting that their typical host halos have masses $\sim$ 2.3 $\times 10^{12}$ \Msolar (based on Mo \& White 2002); while those with SFR $> 100$ \Msolar yr$^{-1}$ are more strongly clustered with $r_0$ $\sim$ 14.1 \mpc$~$ and are hosted by dark matter halos with masses above 2.8$-$4.5 $\times$ $10^{13}$ \Msolar, depending on the actual HOD models.

Figure \ref{fig:r0-dusty} compares our clustering measurement of \textit{sBzK} galaxies to those for other dusty, star-forming systems at similar redshifts. The correlation length we find for \textit{sBzK} galaxies with SFR $> 100$ \Msolar yr$^{-1}$, $14.1 \pm 3.5$ \mpc,~is in broad agreement with that of submillimeter galaxies from recent studies \citep{wei09,wil11,hic12}, which obtained a $r_0$ of $7.7-13.6$ \mpc.
It is also consistent with $r_0 = 12.97^{+4.26}_{-2.64}$ \mpc~ from a measurement of 24 \um-selected ($F_{24} > 0.6$ mJy) dust-obscured galaxies \citep{bro08}, echoing the finding that highly star-forming \textit{BzK}s and brighter dust-obscured galaxies may actually be the same populations \citep{pop08,meg11}. Moreover, the correlation length of our ULIRG-like \textit{sBzK} galaxies is also comparable to that of far-infrared sources detected at 100\um and 160\um with the PACS instrument on $Herschel$ in the GOODS-South field, which have typical correlation lengths of 17.3$-$19 Mpc, or equivalently, 12.2 $-$ 13.3 \mpc, assuming $h = 0.7$ \citep{mag11}.
Despite that the aforementioned populations are selected using different techniques, all these results appear to be converging: at $z \sim 2$, the most rapidly star-forming galaxies are strongly clustered, and highly star-forming BzK's may be linked to and/or overlapping with very dusty systems, based upon the similarities of their correlation lengths.

\subsection{Environment quenching and triggering of SFR?}
In the local Universe, it is known that star formation is a strong function of environment, being less active in dense environments \citep{gom03,kau04}. It is also well established that the star formation rate is correlated with stellar mass at all redshifts \citep{feu05a,noe07,per08}. Since the stellar mass distributions of galaxies in different environments differ, then whether the SFR$-$environment relation is a purely environment effect, or is governed by the stellar mass is still under debate. Recent work at lower redshift ($z < 1$) has suggested that both mass and environment are responsible for shaping the properties of galaxies and that their effects are separable \citep{bam08,pen10,sob11}. The net effect is that in dense environments, or in massive dark-matter halos \footnote{The local density roughly scales with the dark matter halo masses, and therefore for the rest of the discussion we will use environment and dark matter halos interchangeably}, the averaged star formation rate is smaller than that of galaxies located in under-dense regions (or hosted by less-massive halos).

The SFR$-$environment relation can be probed in various ways, including the SFR$-$density, SSFR$-$density, and color$-$density relations. The first two relations can be dervied by measuring the averaged SFR and SSFR as a function of overdensity, while the third relation usually refers to either the relationship between the galaxy color and the overdensity, or the fraction of quiescent galaxies as a function of overdensity. As different probes may result in different results, it is crucial to specify which method is being quoted when interpreting the results as discussed in \citet{pat11}. For instance, at $z\sim 1$, it has been shown that, when considering the populations of both star-forming and quiescent galaxies altogether, the SSFR$-$density and color$-$density relations follow a similar trend as that found locally \citep{coo07,coo08}, while the SFR-density relation is reversed in the sense that the averaged SFR increases with density \citep{elb07,coo08}, primarily due to a population of bright, blue galaxies, as well as dusty LIRGs in overdense environments at this epoch \citep{coo06,elb07,coo08} \footnote{We note that although both \citet{elb07} and \citet{coo08} claimed an inverted SFR$-$density relation at $z \sim 1$, their results regarding the SSFR$-$density do not agree with each other.}. Nevertheless these studies used spectroscopic redshift samples which normally suffer from incomplete sampling of galaxies, leading to large uncertainties in the environment measurements. Recent work by \citet{qua12}, who study the fraction of quiescent galaxies using a photometric redshift sample drawn from the UKIDSS Ultra-Deep Survey (UDS; O. Almaini 2011, in preparation), claims that the color$-$density relation persists out to $z \sim 2$ at all stellar masses, although they caution about the large uncertainties due to the errors in the photometric redshift.

Here we try to address the question of when environment comes to play with a different approach from a local density measurement: we measure the clustering strength as a function of  SSFR among star forming galaxies, and also that for different galaxy types (star-forming vs passive). If galaxies with lower SSFR tend to located in the denser environments, we should see a larger correlation length for galaxies with lower SSFR, based on the assumption that the clustering strength, and hence the halo mass, is strongly correlated with the local density. The advantage of using SSFR instead of SFR is that SSFR measures the star formation efficiency directly, and hence it is easier to interpret the results.

Interestingly, our results suggest that there are two populations separated in the $r_0$$-$SSFR diagram (Figure \ref{fig:r0}d) by SSFR $\sim 2 \times 10^{-9}$ yr$^{-1}$, which corresponds to the main sequence value reported at $z \sim 2$ \citep{dad07,rod11}. For galaxies with SSFR $< 2 \times 10^{-9}$ yr$^{-1}$ (hereafter the low SSFR population), $r_0$ increases rapidly with decreasing SSFR, while $r_0$ increases mildly but significantly with increasing SSFR for galaxies with SSFR above the threshold (hereafter high SSFR population).

\subsubsection{Negative SSFR~$-$~$r_{0}$ relation: environment quenching of star formation}
The anticorrelation between $r_0$ and SSFR for the low SSFR population can be understood as an environmental effect similar to that seen at lower redshifts. For example, using GALEX and SDSS samples at $z < 0.3$, \citet{hei09} found that the galaxy clustering also declines strongly with SSFR. The explanation of those \textit{sBzK} galaxies with very low SSFR in our sample are likely to be that they are falling into the denser environments where their star-formation activities are more effectively suppressed. Since their SSFR values are in between the \textit{sBzK} and \textit{pBzK} galaxies, it suggests that these may be galaxies in transition between the main sequence and the quiescent population. And in fact, their clustering strength is indeed consistent with that of \textit{pBzK} galaxies.

On the other hand, since SSFR anticorrelates with stellar mass for \textit{sBzK} as shown in the right panel of Figure \ref{fig:sfrssfr}, galaxies with lower SSFR tend to have higher \sm$~$ and thus one might expect them to cluster more strongly, providing an alternative explanation for the anticorrelation between $r_0$ and SSFR for low SSFR population. However, such an effect might not be dominant in the redshift range we are probing, as it appears that the stellar mass dependence of SSFR in our sample is weak and can not account for the strong correlation we see in the $r_0$$-$SSFR relation. In order to further test whether the observed $r_0$$-$SSFR relation is due to the combination of $r_0$$-$stellar mass and SSFR-stellar mass correlations, we study the $r_0$$-$SSFR relation in several stellar mass bins wherever we have enough statistics. As shown in Figure \ref{fig:r0-ssfr}, we still see $r_0$ increases with decreasing SSFR for galaxies with SSFR below the critical value $2 \times 10^{-9}$~yr$^{-1}$, even in subsamples divided by stellar mass. We note that this result is robust against the incompleteness due to the missing of relatively quiescent galaxies (red $z-K$ color) in the low stellar mass bins as mentioned in \S 2 since they will only show up in the low SSFR end of the plot if they do exist.

This implies that the $r_0$$-$SSFR relation we see is not purely due to the stellar mass effect. Moreover, not only the \textit{sBzK} galaxies with small SSFR, but the clustering strength of our \textit{pBzK} galaxies is also found to be greater than that of the bulk of \textit{sBzK} galaxies with similar stellar masses (see Figure \ref{fig:r0}), suggesting that quiescent galaxies preferentially reside in denser environments. The negative $r_0$$-$SSFR relation we found for the low SSFR population is analogous to the monotonic decline in the mean overdensity with increasing SSFR that is seen at lower redshifts \citep[e.g. see Figure 9 of][]{coo08}. Furthermore, the higher clustering of \textit{pBzK} galaxies relative to that of \textit{sBzK} galaxies also mirrors the color-density relation found at lower redshifts in which the averaged overdensity is greater for galaxies with redder rest-frame colors \citep{coo06}. In other words, our results imply that some external processes that suppress the star formation activity in dense environments have started as early as $z \sim 2$. On the other hand, the positive correlation between $r_{0}$ and SFR as shown in Figure \ref{fig:r0}c is likely due to the increasing population of  massive star-forming galaxies in denser environment, similar to that found at $z \sim 1$ \citep{elb07,coo08}.

From Figure \ref{fig:r0}, it is also noted that the $r_{0}$$-$SSFR has a steeper slope within the low SSFR range (SSFR $<$ $2 \times 10^{-9}$~yr$^{-1}$), compared to the $r_{0}$$-$\sm~relation, indicating that $r_{0}$ is more sensitive to SSFR than \sm. This trend can be further illustrated in Figure \ref{fig:r0-ssfr}: there exists a clear SSFR dependence of $r_{0}$ when splitting according to \sm, while the variation of $r_{0}$ among different \sm~values is smaller for a fixed SSFR. Similar effect was also observed in a clustering study of the SDSS sample done by \citet{hei09}, who claims that SSFR is a more sensitive probe of the halo mass than \sm. Our result shows that such trend extends out to $z \sim 2$ as well.

\subsubsection{Positive SSFR~$-$~$r_{0}$ relation: environment triggering of star formation}
On the other hand, the trend we see for the high SSFR population is not well-understood.
According to their clustering strength, they have typical halo masses in the range $10^{11}$ to $10^{13}$ \Msolar (see Table \ref{tab:r0}). However, Figure \ref{fig:sfrssfr} shows that galaxies with the largest SSFRs tend to have relatively {\em smaller} stellar masses.  Their stronger clustering therefore seems like a surprising deviation from the main $r_0$ $-$ \sm$~$ relation seen for the $\textit{sBzK}$ population as a whole (Figure \ref{fig:r0}b). Figure \ref{fig:r0-ssfr} shows that this positive correlation
between $r_0$ and SSFR for high SSFR population is seen in all stellar mass bins.

A possible explanation of our result is that the elevated SSFR (``starbursts'') is driven by mechanisms that are associated with more massive halos, even if the stellar masses of the starbursting galaxies are low. Galaxy interactions and mergers, for example, can trigger star formation activity \citep{bar00,lam03,nik04,lin07,ell08}, and they are found to preferentially occur in denser environments \citep{lin10,de11,jia12}. In addition, the large-scale tidal field caused by groups/clusters can also induce starbursts in interacting galaxies \citep{mar08}. The working assumption is thus that the low stellar mass systems with enhanced SSFR are the ones located in denser environments where interactions between galaxies are more common. If this is the case, one may question whether the high clustering amplitude can be attributed to the presence of close neighbors at small scales. However, we emphasize that the close pairs with angular separation less than 0.001 deg are excluded when fitting the clustering amplitude, and therefore the high correlation length found for these high SSFR galaxies is directly linked to their large-scale environments.

Infall shocking is another mechanism which is more effective in massive halos. This may also compress the gas in galaxies, resulting in bursts of star formation. Such environment-driven enhanced star formation activity has also already been found in some cluster studies at $z > 1$ \citep{hil10,tra10}. We note that the enhanced star formation activity in denser environments may not be easily seen in conventional environment studies which look for the quiescent fraction as a function of environment because the red (or blue) fraction may not change significantly even if the averaged star formation efficiency changes.

Our finding concerning a dichotomy in the $r_0$-SSFR relation thus suggests that there are two opposite environmental effects influencing the star formation rate: quenching and triggering. While star formation can be reduced or quenched for galaxies located in denser environments, some other galaxies residing in similar environments can have their star formation rates enhanced instead. Whether the mechanisms responsible for these two effects are related or not is still not clear. One explanation is related to the well-accepted hypothesis that galaxy mergers, commonly found in denser environments, can enhance star formation during the merger process, and then the remnants quickly become red and dead due to the lack of gas which is used up during the starbursting phase, or being blown out by the AGN activity. One way to test this picture is to look for merger signatures of galaxies with elevated or suppressed SSFR. Recent study on the morphologies of starbursting galaxies at $z \sim 2$ already suggests that about 50\% of these sources are associated with interacting and merging galaxies \citep{kar11}. It would be interesting to conduct a similar analysis for galaxies with suppressed SSFR as well.

\section{CONCLUSION}
We have taken advantage of deep $HST$/ACS data and CFHT/WIRCAM NIR data to identify $z \sim 2$ galaxies using \textit{BzK} color selection in the GOODS-N region. We have derived galaxy star-formation rates based on extinction corrected UV luminosity and calculated stellar masses with the $Spitzer$/IRAC 3.6\um and 4.5\um photometry in order to study the clustering properties as a function of $K$-band magnitude, stellar mass (\sm), star formation rate (SFR), and specific star formation rate (SSFR=SFR/\sm). Our main conclusions are as follows:

1. The UV-based SFR, corrected for extinction, scales with stellar mass as SFR $\propto$ \sm$^{\alpha}$, with $\alpha = 0.74\pm0.20$ over the stellar mass range $10^{9}$ \Msolar $<$ \sm $<$ $10^{11}$ \Msolar. The slope of the SSFR$-$\sm$~$is $-0.26\pm0.20$, meaning a weak dependence of star formation rate efficiency on the stellar mass.

2. The clustering amplitude of \textit{sBzK} galaxies is a strong function of $K$-band magnitude and stellar mass, increasing for more massive and brighter galaxies.

3. Highly star-forming galaxies are more strongly clustered than galaxies with low star formation rates, which is in line with the correlation between SFR and stellar mass.

4. For the first time, we find that the correlation length reaches a minimum at SSFR of $2\times 10^{-9}$ yr$^{-1}$, the typical value for the ``main sequence'' of star-forming galaxies at $z \sim 2$. The correlation length is larger for galaxies with both smaller and larger SSFRs. Such a dichotomy holds even at fixed stellar mass. Our results suggest that environment has two effects: quenching and inducing the star formation activities. Stronger clustering for galaxies with relatively low SSFR implies that environment has started playing a role in quenching star formation at $z \sim 2$, while another environment effect, galaxy interactions and mergers, might explain the elevated SSFRs (``starbursts'') in more massive halos (denser environment).

5. Passive galaxies (\textit{pBzK}s) are more strongly clustered than \textit{sBzK} galaxies at a given stellar mass, suggesting that the color$-$density relation is in place at $z \sim 2$. The correlation length $r_0$ of \textit{pBzK} galaxies is measured to be 24.6$\pm$7.0 \mpc, which is larger than that of \textit{sBzK} galaxies but is similar to that of \textit{sBzK} galaxies with the lowest SSFRs.

6. Our results suggest that current Durham semi-analytical models of galaxy formation appear to underestimate the SFR and predict larger scatter in SFR for star-forming galaxies of a given stellar mass. As a result, the predicted $r_0$$-$SFR and $r_0$$-$SSFR relations cannot be directly compared to the observed results, although the $K$ and \sm$~$ dependence of clustering is in better agreement with observations.

One caveat in our analysis, however, lies in the small sample size that prevents us from a more comprehensive HOD analysis. The ongoing $Spitzer$ Extended Deep Survey (SEDS; PI: G. Fazio), combined with existing multi-wavelength data in several extra-galactic fields, will provide larger samples with better determined photometric redshifts and stellar mass measurements. This survey will thus allow improved modeling of the relationships between  halo mass and the galaxy properties through the combined clustering and abundance analysis.

\begin{deluxetable}{lcc}
\tabletypesize{\scriptsize}
\tablewidth{0pt}
\tablecaption{The values of $a_{1}$ and $a_{2}$ for the empirical formula (Eq. \ref{eq:sm}) to compute the stellar mass.\label{tab:smfit}}
\tablehead{
    \colhead{Redshift} &
    \colhead{$a_{1}$} &
    \colhead{$a_{2}$}
}

\startdata
$1.0 < z < 1.5$ &19.65 &-0.084\\
$1.5 < z < 2.0$ &20.21 &0.069\\
$2.0 < z < 2.5$ &20.79 &0.256\\
$2.5 < z < 3.0$ &20.80 &0.890

\enddata

\end{deluxetable}


\begin{deluxetable*}{lcccccccc}

\tabletypesize{\scriptsize}
\tablewidth{0pt}

\tablecaption{Clustering properties and corresponding dark matter halo masses of \textit{sBzK} and \textit{pBzK} galaxies in GOODS-N.\label{tab:r0}}
\tablehead{
    \colhead{Sample Cut} &
    \colhead{$\overline{z}$} &
    \colhead{number} &
    \colhead{\Aw$^{(a)}$} &
    \colhead{$r_0$$^{(b)}$} &
    \colhead{bias$^{(c)}$} &
    \colhead{$M_{DM}^{1 (d)}$} &
    \colhead{$M_{DM}^{2 (e)}$} &
    \colhead{$M_{DM}^{3 (f)}$}
}

\startdata
$21.0 < K_{\rm s} < 21.5$           & 2.21 & 86 & $0.014270 \pm 0.004534$ & $17.4 \pm 4.9$ & $5.2 \pm 0.8$ & $(7.2 \pm 2.9) \times 10^{13}$ & $(5.0 \pm 2.3) \times 10^{13}$ & $(4.9 \pm 2.3) \times 10^{13}$ \\
$21.5 < K_{\rm s} < 22.0$           & 2.13 & 183 & $0.005461 \pm 0.001084$ & $10.3 \pm 2.5$ & $3.3 \pm 0.3$ & $(1.9 \pm 0.6) \times 10^{13}$ & $(1.1 \pm 0.4) \times 10^{13}$ & $(1.0 \pm 0.4) \times 10^{13}$ \\
$22.0 < K_{\rm s} < 22.5$           & 2.15 & 295 & $0.003008 \pm 0.000658$ & $7.4 \pm 1.9$ & $2.4 \pm 0.3$ & $(7.1 \pm 2.8) \times 10^{12}$ & $(3.4 \pm 1.5) \times 10^{12}$ & $(3.0 \pm 1.5) \times 10^{12}$ \\
$22.5 < K_{\rm s} < 23.0$           & 2.24 & 441 & $0.002025 \pm 0.000491$ & $5.8 \pm 1.5$ & $2.0 \pm 0.2$ & $(3.2 \pm 1.6) \times 10^{12}$ & $(1.3 \pm 0.8) \times 10^{12}$ & $(1.1 \pm 0.7) \times 10^{12}$ \\
$23.0 < K_{\rm s} < 23.5$           & 2.19 & 702 & $0.002514 \pm 0.000256$ & $6.6 \pm 1.5$ & $2.2 \pm 0.1$ & $(5.0 \pm 1.0) \times 10^{12}$ & $(2.2 \pm 0.5) \times 10^{12}$ & $(1.9 \pm 0.5) \times 10^{12}$ \\
$23.5 < K_{\rm s} < 24.0$           & 2.12 & 1029 & $0.001151 \pm 0.000165$ & $4.3 \pm 1.0$ & $1.5 \pm 0.1$ & $(9.3 \pm 3.4) \times 10^{11}$ & $(3.4 \pm 1.4) \times 10^{11}$ & $(2.1 \pm 1.1) \times 10^{11}$ \\
$9.0 < log(M_{*}$/\Msolar) $< 9.5$          & 1.78 & 1688 & $0.001157 \pm 0.000104$ & $4.5 \pm 1.0$ & $1.5 \pm 0.1$ & $(1.3 \pm 0.3) \times 10^{12}$ & $(4.7 \pm 1.2) \times 10^{11}$ & $(2.9 \pm 0.9) \times 10^{11}$ \\
$9.5 < log(M_{*}$/\Msolar) $< 10.0$         & 2.13 & 1201 & $0.001817 \pm 0.000135$ & $5.6 \pm 1.2$ & $1.9 \pm 0.1$ & $(2.7 \pm 0.4) \times 10^{12}$ & $(1.1 \pm 0.2) \times 10^{12}$ & $(8.9 \pm 1.7) \times 10^{11}$ \\
$10.0 < log(M_{*}$/\Msolar) $< 10.5$                & 2.23 & 732 & $0.002291 \pm 0.000255$ & $6.3 \pm 1.4$ & $2.1 \pm 0.1$ & $(4.1 \pm 0.9) \times 10^{12}$ & $(1.8 \pm 0.5) \times 10^{12}$ & $(1.5 \pm 0.4) \times 10^{12}$ \\
$10.5 < log(M_{*}$/\Msolar) $< 11.0$                & 2.26 & 340 & $0.005888 \pm 0.000636$ & $10.6 \pm 2.4$ & $3.3 \pm 0.2$ & $(2.1 \pm 0.4) \times 10^{13}$ & $(1.2 \pm 0.2) \times 10^{13}$ & $(1.1 \pm 0.2) \times 10^{13}$ \\
$log(M_{*}$/\Msolar) $> 11.0$               & 2.27 & 107 & $0.009662 \pm 0.002961$ & $13.9 \pm 3.9$ & $4.3 \pm 0.7$ & $(4.3 \pm 1.8) \times 10^{13}$ & $(2.7 \pm 1.3) \times 10^{13}$ & $(2.6 \pm 1.3) \times 10^{13}$ \\
1.0 $<$ SFR/(\Msolar yr$^{-1}$) $<$ 5.0             & 1.84 & 851 & $0.001527 \pm 0.000186$ & $5.3 \pm 1.2$ & $1.7 \pm 0.1$ & $(2.3 \pm 0.6) \times 10^{12}$ & $(9.1 \pm 2.7) \times 10^{11}$ & $(6.8 \pm 2.5) \times 10^{11}$ \\
5.0 $<$ SFR/(\Msolar yr$^{-1}$) $<$ 10              & 2.06 & 1013 & $0.001392 \pm 0.000162$ & $4.9 \pm 1.1$ & $1.7 \pm 0.1$ & $(1.5 \pm 0.4) \times 10^{12}$ & $(5.9 \pm 1.8) \times 10^{11}$ & $(4.3 \pm 1.6) \times 10^{11}$ \\
10 $<$ SFR/(\Msolar yr$^{-1}$) $<$ 30               & 2.13 & 1243 & $0.001814 \pm 0.000133$ & $5.6 \pm 1.2$ & $1.9 \pm 0.1$ & $(2.7 \pm 0.4) \times 10^{12}$ & $(1.1 \pm 0.2) \times 10^{12}$ & $(8.9 \pm 1.8) \times 10^{11}$ \\
30 $<$ SFR/(\Msolar yr$^{-1}$) $<$ 60               & 2.27 & 484 & $0.002571 \pm 0.000437$ & $6.6 \pm 1.6$ & $2.2 \pm 0.2$ & $(5.0 \pm 1.6) \times 10^{12}$ & $(2.3 \pm 0.8) \times 10^{12}$ & $(2.0 \pm 0.8) \times 10^{12}$ \\
60 $<$ SFR/(\Msolar yr$^{-1}$) $<$ 100              & 2.33 & 203 & $0.005680 \pm 0.001059$ & $10.2 \pm 2.5$ & $3.3 \pm 0.3$ & $(1.9 \pm 0.6) \times 10^{13}$ & $(1.1 \pm 0.4) \times 10^{13}$ & $(1.0 \pm 0.4) \times 10^{13}$ \\
SFR/(\Msolar yr$^{-1}$)  $>$ 100.0          & 2.21 & 218 & $0.009798 \pm 0.002099$ & $14.1 \pm 3.5$ & $4.4 \pm 0.5$ & $(4.5 \pm 1.3) \times 10^{13}$ & $(2.8 \pm 0.9) \times 10^{13}$ & $(2.8 \pm 0.9) \times 10^{13}$ \\
$2\times10^{-10}$  $<$ SSFR/yr$^{-1}$ $<$ $4.5\times10^{-10}$       & 2.33 & 105 & $0.018360 \pm 0.002629$ & $19.6 \pm 4.6$ & $5.8 \pm 0.4$ & $(9.6 \pm 1.6) \times 10^{13}$ & $(6.8 \pm 1.4) \times 10^{13}$ & $(6.8 \pm 1.4) \times 10^{13}$ \\
$4.5\times10^{-10}$ $<$ SSFR/yr$^{-1}$ < $7.5\times10^{-10}$        & 2.22 & 146 & $0.005615 \pm 0.001473$ & $10.3 \pm 2.7$ & $3.3 \pm 0.4$ & $(1.9 \pm 0.8) \times 10^{13}$ & $(1.1 \pm 0.5) \times 10^{13}$ & $(1.0 \pm 0.5) \times 10^{13}$ \\
$7.5\times10^{-10}$  $<$ SSFR/yr$^{-1}$ $<$ $1.5\times10^{-9}$      & 2.13 & 708 & $0.001666 \pm 0.000267$ & $5.3 \pm 1.3$ & $1.8 \pm 0.1$ & $(2.2 \pm 0.8) \times 10^{12}$ & $(8.9 \pm 3.6) \times 10^{11}$ & $(6.9 \pm 3.2) \times 10^{11}$ \\
$1.5\times10^{-9}$ $<$ SSFR/yr$^{-1}$ $<$ $3.5\times10^{-9}$                & 2.13 & 1883 & $0.001141 \pm 0.000090$ & $4.3 \pm 1.0$ & $1.5 \pm 0.1$ & $(9.1 \pm 1.8) \times 10^{11}$ & $(3.3 \pm 0.7) \times 10^{11}$ & $(2.1 \pm 0.6) \times 10^{11}$ \\
$3.5\times10^{-9}$ $<$ SSFR/yr$^{-1}$ $<$ $7\times10^{-9}$          & 1.79 & 847 & $0.003525 \pm 0.000308$ & $8.4 \pm 1.9$ & $2.6 \pm 0.1$ & $(1.3 \pm 0.2) \times 10^{13}$ & $(6.2 \pm 1.1) \times 10^{12}$ & $(5.6 \pm 1.0) \times 10^{12}$ \\
SSFR/yr$^{-1}$  $>$ $7\times10^{-9}$                & 2.38 & 301 & $0.005726 \pm 0.000639$ & $10.2 \pm 2.3$ & $3.3 \pm 0.2$ & $(1.9 \pm 0.3) \times 10^{13}$ & $(1.0 \pm 0.2) \times 10^{13}$ & $(1.0 \pm 0.2) \times 10^{13}$ \\
\hline
pBzK's      & 1.99 & 44 & $0.025320 \pm 0.008292$ & $24.6 \pm 7.0$ & $7.1 \pm 1.2$ & $(1.5 \pm 0.6) \times 10^{14}$ & $(1.2 \pm 0.5) \times 10^{14}$ & $(1.2 \pm 0.5) \times 10^{14}$
\enddata

\tablecomments{All values listed above are for the \textit{sBzK} galaxies except for the last row, which is for the \textit{pBzK} galaxies. $^{(a)}$ Correlation amplitude at 1 degree; $^{(b)}$ Correlation length in \mpc ; $^{(c)}$ Bias ; $^{(d)}$ Halo mass in \Msolar$~$inferred directly from the \citet{mo02} formalism for the value of $r_0$ ; $^{(e)}$ Minimum halo mass in \Msolar$~$assuming $r_0$ is the effective clustering strength averaged over halos above a certain threshold; $^{(f)}$ Minimum halo mass in \Msolar$~$assuming that $r_0$ is the effective clustering strength averaged over halos above a certain threshold and weighted by the applied HOD, as described in the text.}

\end{deluxetable*}

\acknowledgments

We thank the anonymous referee for the thorough reading and constructive report, which improved the quality of this work. We thank M. Kriek for her permission of using the SED fitting code 'FAST', A. Coil, M. Cooper, J. Coupon, I. Smail, and R. Hickox for their helpful discussions, and M. Brodwin for the assistance on the photometric redshift measurement during the earlier phase of this work. We also thank Loic Albert and other CFHT staffs who have provided valuable suggestions for the WIRCAM data reduction. The work is supported by the National Science Council of Taiwan
under the grant NSC99-2112-M-001-003-MY3. H.-Y. Jian acknowledges the support of NSC grant NSC97-2628-M-002-008-MY3 and D. C. Koo of NSF grant AST 08-08133. The PhD work of A. I. Merson is funded by the Science \& Technologies Facilities Council (STFC). This work is also based in part on observations obtained with WIRCam, a joint project of CFHT, Taiwan, Korea, Canada, France, at the Canada-France-Hawaii Telescope (CFHT) which is operated by the National Research Council (NRC) of Canada, the Institute National des Sciences de l'Univers of the Centre National de la Recherche Scientifique of France, and the University of Hawaii. Access to the CFHT was made possible by the Ministry of Education and the National Science Council of Taiwan as part of the Cosmology and Particle Astrophysics (CosPA) initiative. We close with thanks to Hawaiian people for the use of their sacred mountain.

\end{document}